\documentclass[superscriptaddress,aps,prl,floats,onecolumn,twoside,floatfix]{revtex4}
\usepackage{epsfig}
\usepackage{amsmath}
\usepackage{amssymb}
\usepackage[english]{babel}

\def\be{\begin{equation}}
\def\ee{\end{equation}}
\def\bea{\begin{eqnarray}}
\def\eea{\end{eqnarray}}

\def\<{\langle}
\def\>{\rangle}
\def\dl{\langle\langle}
\def\dr{\rangle\rangle}
\def\~{\tilde}
\def\s{\sigma}
\def\l{\lambda}

\def\b{\beta}

\def\o{\omega}
\def\t{\tau}

\newcommand{\E}{\Bbb E}

\newcommand{\av}[1]{\mbox{{\rm Av}}\left(#1\right)}

\newenvironment{proof}{Proof:}{\hfill$\square$\vskip.5cm}
\newcommand{\beq}{\begin{eqnarray}}
\newcommand{\eeq}{\end{eqnarray}}

\begin{document}
\title{Stability of the Spin Glass Phase under Perturbations}
\author{Pierluigi Contucci}
\affiliation{Universit\`{a} di Bologna, Piazza di Porta S.Donato 5, 40127 Bologna, Italy}

\author{Cristian Giardin\`a}
\affiliation{Universit\`a di Modena e Reggio E., viale A. Allegri, 9 - 42121 Reggio Emilia , Italy}

\author{Claudio Giberti}
\affiliation{ Universit\`a di Modena e Reggio E., via G.
Amendola 2 -Pad. Morselli- 42122 Reggio Emilia, Italy}

\begin{abstract}
We introduce and prove a novel linear response stability theory for spin glasses.
The new stability under suitable perturbation of the equilibrium state
implies the whole set of structural identities that characterize the spin glass phase.
\end{abstract}

\maketitle

The linear response stability under small perturbations is a basic method in statistical physics \cite{K},\cite{LL}. By suitably perturbing the energy of a system it allows to obtain relations (structural identities) among physical quantities by simply computing a derivative. Those identities identify the order parameters of the theory thus reducing its apriori degrees of freedom. The mean field theory for the ferromagnetic phase in which the only order parameter is the magnetization is a celebrated example of such an approach.

The purpose of this paper is to introduce a novel linear response stability for spin glasses. Their low temperature phase is still subject of intense investigations and debates in physics. The importance of the spin glass problem is not only related to the condensed matter original formulation but also to the numerous applications that its methods and techniques have brought into other disciplines \cite{MPV},\cite{FH},\cite{MM}. At the core of the problem there is the much awaited solution of the debate between droplet \cite{FHu} and mean field \cite{MPV} pictures, as well as the trivial versus hierarchical (ultrametric) organization of pure states in the low temperature phase. If such a debate is basically settled in the mean field approximation the physics community hasn't reached yet any consensus on the relevance of the ultrametricity for the short range finite dimensional models (see nevertheless the numerical results in \cite{CGGPV}).

In a former linear stability theory for the spin glass phase (stochastic stability) introduced in \cite{AC} it was shown how to derive an infinite family of polynomial structural identities (see \cite{Ba} at equilibrium and \cite{FMPP} out of equilibrium). On the other hand a larger set of identities, known as Ghirlanda-Guerra identities \cite{GhiGu}, has been obtained by the control of the energy fluctuations with respect to the equilibrium state first in the mean-field approximation and then also for short-range models \cite{CG1},\cite{CG2}.

From the linear response stability that we introduce and prove in this paper we obtain the entire set of Ghirlanda-Guerra identities of the spin glass theory. By consequence their derivation requires only a simple computation of a derivative, expressing the stability under a suitable perturbation of the equilibrium state. The new perturbation reproduces both thermal and disorder fluctuations and is able to bypass the refined fluctuation theory bounds of the quenched measure used by the mathematical physics community.

The Gibbs-Boltzmann state $\omega_{\b,N}$
of a statistical mechanics system of $N$ interacting spins $\s=(\s_1,...,\s_N)$,
with Hamiltonian $H(\sigma)$ at inverse temperature $\b$, admits the classical probabilistic interpretation
as the {\it deformation} of the uniform measure over spin configurations $\mu_N$:
\be
\o_{\b,N}(f) \; = \; \frac{\mu_N(fe^{-\beta H})}{\mu_N(e^{-\beta H})} \; ,
\ee
with
\be
\mu_N(f) \; = \; \frac{1}{2^N}\sum_{\sigma}f(\sigma) \; ,
\ee
and $f$ a smooth bounded function of the spin configurations.
Such a deformed state $\omega_{\b,N}$ fulfills a remarkable
{\it stability property} with respect to further small deformations ({\it perturbations}):
considering the Hamiltonian per particle
\be
h(\s)=\frac{H(\s)}{N}
\ee
and the perturbation with parameter $\l$ defined as
\be
\label{dp}
\o_{\b,N}^{(\l)}(f) \; = \; \frac{\o_{\b,N}(fe^{-\l h})}{\o_{\b,N}(e^{-\l h})}
\ee
the Gibbs-Boltzmann measure is stable, i.e.\;$\l$-independent, in the thermodynamic limit
$N \rightarrow \infty$.
In fact one can observe that the perturbation amounts to a small
temperature shift:
\be\label{tsd}
\o_{\b,N}^{(\l)}(f) \; = \; \frac{\mu_N(fe^{-\beta H - \l h})}{\mu_N(e^{-\beta H - \l h})} \; = \; \o_{\b+\frac{\l}{N},N}(f)
\ee
which implies that it has a vanishing effect in the large volume limit apart form isolated singularities,
possibly related to phase transitions.
More precisely one can prove the stability as follows: since for all $\b$ intervals and all values of $\l$ one has, thanks to (\ref{tsd}),
\beq
&&\int_{\b_0}^{\b_1}\frac{d \o_{\b,N}^{(\l)}(f)}{d\l}d\b \; =\; \frac{1}{N} \int_{\b_0}^{\b_1} \frac{d \o_{\b,N}^{(\l)}(f)}{d\b}d\b\\ &=&
\frac{\o_{\b_1,N}^{(\l)}(f)-\o_{\b_0,N}^{(\l)}(f)}{N} \;
\eeq
one obtains:
\be
\lim_{N\to \infty}\int_{\b_0}^{\b_1} \frac{d \o_{\b,N}^{(\l)}(f)}{d\l}d\b \; = \; 0 \quad \forall \;\;\l , \quad \forall \;\; [\beta_0,\beta_1] \; .
\ee
As a consequence, computing the derivative at $\l=0$, one has
\be
\label{fluttua}
\lim_{N\to \infty}\int_{\b_0}^{\b_1} \left[ \o_{\b,N}(fh) - \o_{\b,N}(f)\o_{\b,N}(h) \right] d\b \; = \; 0  \; .
\ee
For the special case $f=h$ the previous formula implies that
the Hamiltonian per particle converges to a constant for
large volumes
with respect to the Gibbs measure,
at least in $\beta-$integral average.
Higher order derivatives with respect to $\lambda$ of
$\o_{\b,N}^{(\l)}(f)$ (i.e. cumulants of $h$)
are then enforced to vanish since
cumulants are homogeneous polynomials of
the constant values of $h$ with coefficients whose
sum is zero.

Formula (\ref{fluttua}) has interesting consequences. It says, for instance, that
the order parameter (i.e. the magnetization) for a mean field ferromagnetic Hamiltonian
has a trivial distribution. In the Curie-Weiss model at zero magnetic field,
for which the Hamiltonian per particle is the square magnetization, the previous
identity implies (by choosing $f=h$) that in the thermodynamic limit,
\be
\label{sopra}
\o_{\b}(\s_1\s_2\s_3\s_4) \; = \; \o_{\b}(\s_1\s_2)^2
\ee
in $\beta-$average \cite{CGI}. One can indeed prove that
(\ref{sopra}) holds for all $\beta$ using the methods
developed in \cite{EN}.
The choice $f=h^n$, or equivalently higher
order derivatives in $\l$ of the perturbed state,
gives the well known factorization property of the $2n$-point function
as an $n$-th power of the $2$-point function.

In a disordered system defined by a centered
Gaussian Hamiltonian ${\cal H}(\s)$ of covariance
(generalized overlap)
\be
\av{{\cal H}(\s){\cal H}(\t)} \; = \; N c_N(\s,\t)
\ee
the equilibrium measure is the quenched average of the random Boltzmann-Gibbs
state $\o_{\b,N}$: for a bounded random function $f$ it is defined by
\be
\< f \>_{\b,N} \; = \; \av{\o_{\b,N}(f)} \; .
\ee

The thermodynamic properties of the system are expressed in terms of
a set of random variables $ \{c_{i,j}\}$
related to the quenched expectation of the covariance entries.
Namely, considering the Boltzmann-Gibbs product state
$\Omega_{\b,N} = \o_{\b,N} \times \o_{\b,N}$, one defines
the random variables  $c_{i,j}$ and their joint distribution by:
\be
\E_{\b,N}(c_{i,j}) = \av{ \Omega_{\b,N}(c(\s^{(i)},\s^{(j)}))}\;.
\ee
In \cite{AC} it was identified a {\it stochastic stability} property of the quenched state,
i.e. an invariance with respect to the stochastic perturbation:
\be\label{ssp}
\< f \>_{\b,N}^{(\l)} \; = \; \av{\frac{\o_{\b,N}(fe^{\sqrt{\lambda}{\cal K}})}{\o_{\b,N}(e^{\sqrt{\lambda}{\cal K}})} } \; ,
\ee
where ${\cal K}(\s)$ is a random field (independent from the Hamiltonian) whose
covariance is $c_N(\s,\t)$, and it was shown that the stochastic perturbation is equivalent to
a temperature shift
\be
\< f \>_{\b,N}^{(\l)} \; = \; \< f \>_{\sqrt{\b^2+\frac{\l}{N}},N}
\; ,
\ee
from which stability follows \cite{CG1}
\be
\lim_{N\to \infty}\int_{\b_0}^{\b_1} \frac{d \<f\>_{\b,N}^{(\l)}}{d\l}d\b \; = \; 0 \quad \forall \;\;\l , \quad \forall \;\; [\beta_0,\beta_1] \; .\ee
By consequence at $\l=0$ one obtains:
\be\label{poly}
\lim_{N\to \infty}\int_{\b_0}^{\b_1} \av{ \o_{\b,N}(fh) - \o_{\b,N}(f)\o_{\b,N}(h) } d\b \; = \; 0  \; .
\ee
The previous formula implies (taking $f=h$ and integrating by parts)
\be\label{ac1}
\lim_{N\to \infty}\int_{\b_0}^{\b_1}\E_{\b,N}(  c^2_{1,2} - 4 c_{1,2} c_{2,3} + 3 c_{1,2} c_{3,4} ) \; d\b \; = \; 0 \; .
\ee
We stress the fact that the previous identity holds for a general Gaussian Hamiltonian, both mean field
or short range, in terms of its own covariance.
For $f=h^n$ one can see \cite{CG1} that the identities that can be derived from (\ref{poly})
are, like the (\ref{ac1}), zero average polynomials in the $c_{i,j}$ with respect to the
quenched measure. See also  \cite{Ba} for an alternative derivation.

In \cite{Gu} it was introduced a method, based on bounds for the energy fluctuations,
which leads to the set of Ghirlanda-Guerra identities \cite{GhiGu,CG2,Bov,TaL}; the lowest order
are for instance:
\be
\E_{\b,N}( c_{1,2} c_{2,3}) \;=\; \frac12\,\E_{\b,N}( c_{1,2}^2)+ \frac12\,\E_{\b,N}( c_{1,2})^2
\ee
\be
\E_{\b,N}( {c}_{1,2} c_{3,4}) \;=\; \frac13\,\E_{\b,N}( c_{1,2}^2) + \frac23\,\E_{\b,N}( c_{1,2})^2\;.
\ee
Unlike the set of identities that can be derived from stochastic stability, these also include
non-linear terms of the overlap expectations.  In recent times it was shown that an invariance
under reshuffling introduced in the framework of competing particle systems \cite{ArAi} implies the whole
set of Ghirlanda-Guerra identities \cite{A}.

To this purpose we introduce and prove here a novel stability property for the spin glass quenched state.
We show that from such a stability property the whole
set of Ghirlanda-Guerra relations can be derived.


We define the perturbation of quenched state as
\be
\label{gs}
\dl f \dr^{(\l)}_{\b,N} \; = \; \frac{\av{\o_{\b,N}(f e^{-\l h})}}{\av{\o_{\b,N}(e^{-\l h})}}\;.
\ee

We observe that this new perturbation is the analog, for the quenched measure
of a random Hamiltonian, of the standard perturbation (\ref{dp}) introduced
for deterministic systems with respect to the Boltzmann-Gibbs measure.
On the other hand we notice that while the stochastic stability perturbation (\ref{ssp}),
as much as the standard perturbation for deterministic system, amounts
to a small temperature shift, the newly introduced perturbation
cannot be reduced to just a small temperature change {(see the remark after formula (\ref{fsr}))}. More precisely
the explicit expression of (\ref{gs}) reads
\be
\dl f \dr^{(\l)}_{\b,N} \; = \; \frac
{\av{ \frac{\sum_{\s}f(\s)e^{-(\b+\l/N)H(\s)}}{\sum_{\s} e^{-\b H(\s)}       } }}
{\av{ \frac{\sum_{\s}e^{-(\b+\l/N)H(\s)}}{\sum_{\s} e^{-\b H(\s)}       } }}
\ee
where it clearly appears that only the numerator of the random Boltzmann-Gibbs
state is affected by the change.

Our main result is summarized by the following
\medskip
\par\noindent
{\bf Proposition}
{\em
With the definition given above, the quenched state of a
Gaussian spin glass is stable under the deformation
(\ref{gs}), i.e.
\be\label{gsp}
\lim_{N\to \infty}\int_{\b_0}^{\b_1} \left.\frac{d \dl f\dr_{\b,N}^{(\l)}}{d\l}\right|_{\l=0} d\b \; = \; 0 \; .
\ee
Moreover the property (\ref{gsp}) implies the whole set of the Ghirlanda-Guerra identities:
for a bounded $f$ function of the generalized overlaps $\{c_{i,j}\}$ (with $i,j \in \{1,...,n\}$)
{the quantity $\delta_N(\beta)$ defined by
\begin{eqnarray}
\label{ggm}
\E_{\b,N}(f \,c_{1,n+1})&=&\frac{1}{n} \E_{\b,N}(f)\,\E_{\b,N}(c_{1,2})\nonumber \\
&+& \frac{1}{n} \sum_{j=2}^n \E_{\b,N}(f c_{1,j}) + \delta_N(\beta)
\end{eqnarray}
goes to zero in $\beta$-average and for large volumes.}}

\begin{proof}
A simple calculation shows that
\be
\left.\frac{d \dl f\dr_{\b,N}^{(\l)}}{d\l}\right|_{\l=0}  \; = \;  \<fh\>_{\b,N}-\<f\>_{\b,N}\<h\>_{\b,N} \; .
\ee
The right hand side can be decomposed into two terms which can be identified as the thermal
and the disorder correlations:
\begin{eqnarray}
\hspace{-0.5cm}&&\left.\frac{d \dl f\dr_{\b,N}^{(\l)}}{d\l}\right|_{\l=0}\hspace{-0.5cm}
= \av{\o_{\b,N}(fh)-\o_{\b,N}(f)\o_{\b,N}(h)}+ \nonumber \\
& + &\hspace{-0.2truecm} \av{\o_{\b,N}(f)\o_{\b,N}(h)}-\av{\o_{\b,N}(f)}\av{\o_{\b,N}(h)}.\nonumber \\
& &
\end{eqnarray}
In \cite{CG2} the two previous terms were proved to converge to zero in $\beta$ average
and using integration Gaussian by parts it was shown how they imply formula (\ref{ggm}).
\end{proof}
\noindent
{\bf Remarks}

{\em
i) It is interesting to notice that the new stability property introduced in this paper as well as
those introduced in the past admit a simple formulation in terms of cumulant generating function.
Defining that function for the quenched state as
\be\label{qgf}
\psi_{\beta,N}(\lambda) \; = \; \ln \av{\frac{Z_{\b+\l/N}}{Z_{\b}}} \; = \; \ln \<e^{\l h }\>_{\beta,N}
\ee
the (\ref{gsp}) is equivalent, {by the boundedness of $f$ and the Schwarz inequality
\be\label{si}
|\<fh\>-\<f\>\<h\>| \le \sqrt{\<f^2\>
}\sqrt{\<h^2\>-\<h\>^2} \; ,
\ee
}
to the property of asymptotic flatness at the origin
\be\label{af}
\lim_{N\to \infty}\int_{\b_0}^{\b_1} \left.\frac{d^2  \psi_{\b,N}{(\l)}}{d\l^2}\right|_{\l=0} d\b \; = \; 0 \; .
\ee
In particular defining the generating function of thermal fluctuations as
\be\label{ssgf}
{\bar\psi}_{\beta,N}(\lambda) \; = \; \av{ \ln \o_{\beta,N}(e^{\l h })}
\ee
and the generating function of disorder fluctuations as
\be\label{dgf}
{\tilde\psi}_{\beta,N}(\lambda) \; = \; \ln \av{e^{\l \o_{\beta,N}(h)}}
\ee
one has
\be\label{fsr}
{\left.\frac{d^2  \psi_{\b,N}{(\l)}}{d\l^2}\right|_{\l=0} \; = \; \left.\frac{d^2  {\bar\psi}_{\b,N}{(\l)}}{d\l^2}\right|_{\l=0} + \left.\frac{d^2  {\tilde\psi}_{\b,N}{(\l)}}{d\l^2}\right|_{\l=0} }\; ,
\ee
{which shows that the new perturbation can be seen as a sum of thermal and disorder perturbations.

ii) The statement of the Proposition can be made to hold point-wise in $\beta$ under suitable technical
assumptions, namely the differentiability of the free energy. This indeed has been shown for the standard
stochastic stability in \cite{ArCh} and for the Ghirlanda-Guerra identities in \cite{Panz}.
}
}

%
%
%

\vspace{0.5truecm}
The results shown in this paper provides a straightforward method to obtain the structural identities of the spin glass phase known as Ghirlanda-Guerra identities by a simple computation of a derivative and a Gaussian integration by parts. This provides a new interpretation, using a stability argument, of the vanishing fluctuation property from which they were originally derived \cite{GhiGu}.
The relevance of the stability properties and of the Ghirlanda-Guerra identities has been shown in the work \cite{ArAi} and \cite{Pan} where, under the hypothesis of discreteness of the overlap distribution it was proved, respectively, that competing particle systems satisfying invariance under reshuffling or spin systems satisfying Ghirlanda-Guerra identities do fulfill the hierarchical structure (ultrametricity) originally introduced in the Parisi work for the mean field spin glass \cite{MPV}.
The present work provides a further bridge between those two approaches, whose mutual relation has still to be fully clarified \cite{Tala}, suggesting that the invariance under reshuffling is well represented by our newly introduced stability under perturbation.

\vspace{0.5truecm}
{\bf Acknowledgments.} 
We thank STRATEGIC RESEARCH GRANT (University
of Bologna) and INTERNATIONAL RESEARCH PROJECTS (Fondazione
Cassa di Risparmio, Modena) for partial financial support.

\end{document}